\documentclass[twocolumn,aps,prb]{revtex4-1}

\usepackage[utf8x]{inputenc}
\usepackage{amsmath,amssymb,amsthm}
\usepackage{mathtools}
\usepackage{xspace}
\usepackage[usenames,dvipsnames]{xcolor}
\usepackage[colorlinks,hyperindex,breaklinks]{hyperref}
\usepackage{braket}
\usepackage{hyphenat}
\usepackage{bbold}
\usepackage{enumerate}
\usepackage{tabularx}
\usepackage[verbose]{placeins}
\usepackage[hang,flushmargin]{footmisc}
\usepackage{comment}

\usepackage[ngerman,british]{babel}

\renewcommand{\vec}[1]{\mathbf{#1}}

\begin{document}

\title{Calculating linear response functions for finite temperatures 
on the basis of the alloy analogy model}

\author{H. Ebert, S. Mankovsky, K. Chadova, S. Polesya, 
J. Min\'{a}r, and D. K\"odderitzsch}

\affiliation{Department Chemie/Phys.\ Chemie, 
            Ludwig-Maximilians-Universit\"at
           M\"unchen, Butenandtstrasse 5-13, D-81377 M\"unchen, Germany}

\date{\today}


\begin{abstract}
A scheme is presented that is based on the alloy analogy model and
allows to account for thermal lattice vibrations as well as spin
fluctuations when calculating response quantities in solids.  
Various models to deal with spin fluctuations are discussed
concerning their impact on the resulting temperature dependent
magnetic moment, longitudinal conductivity and Gilbert
damping parameter. 
It is demonstrated that using the Monte Carlo (MC) spin
configuration as an input, the alloy analogy model is capable to
reproduce results of MC simulations on the average magnetic moment
within all spin fluctuation models under discussion. 
On the other hand, response quantities 
are much more sensitive to the spin fluctuation model.
Separate calculations accounting for either the
thermal effect due to lattice vibrations or spin fluctuations show
their comparable contributions to the electrical
conductivity and Gilbert damping. However, comparison to
results accounting for both thermal effects demonstrate 
violation of Matthiessen’s rule, showing the non-additive effect
of lattice vibrations and spin fluctuations.
The results obtained for bcc Fe and fcc Ni are compared with
the experimental data, showing rather good agreement for the temperature
dependent electrical conductivity and Gilbert damping parameter. 
\end{abstract}


\maketitle

\section{Introduction}

Finite temperature has often a very crucial influence on the 
response properties of a solid. A prominent example for this 
is the electrical resistivity of perfect non-magnetic metals 
and ordered compounds that only take a non-zero value with a
 characteristic temperature ($T$) dependence due to thermal lattice
 vibrations. While the Holstein transport equation \cite{Hol64,Mah00} provides
 a sound basis for corresponding calculations numerical work in this 
field has been done so far either on a model level or for
 simplified situations.\cite{All71,TW77,Gri76,MH83} 
In practice often the Boltzmann-formalism 
is adopted using the constant relaxation time ($\tau$) approximation.
 This is a very popular approach in particular when dealing with the 
Seebeck effect, as in this case $\tau$ drops out.\cite{OYAA09,SY13}
The constant relaxation time approximation has also been used 
extensively when dealing with the Gilbert damping 
parameter $\alpha$.\cite{SF05,Kam07,GIS08} Within the description of 
Kambersky \cite{Kam76,Kam07}
the conductivity- and resistivity-like 
intra- and inter-band contributions to $\alpha$ show a 
different dependency on $\tau$ leading 
typically to a minimum for 
$\alpha(\tau)$ or equivalently for $\alpha(T)$.\cite{Kam07,GIS08} 
A scheme
 to deal with the temperature dependent resistivity that 
is formally much more satisfying than the constant relaxation
 time  approximation is achieved by combining the 
Boltzmann-formalism with a detailed calculation 
of the phonon properties. 
As was shown by various authors,\cite{ABK+86,SS96,XV13,XV14}
this parameter-free approach leads for non-magnetic metals in
 general to a very good agreement with experimental data.

As an alternative to this approach, thermal 
lattice vibrations have also been accounted for 
within various studies by quasi-static lattice 
displacements leading to thermally induced structural 
disorder in the system. This point of view provides the
 basis for the use of the alloy analogy, i.e.\ for 
the use of
 techniques to deal with substitutional chemical disorder 
also when dealing with temperature 
dependent 
quasi-static random  lattice displacements.
 An example for this are
 investigations on the temperature dependence of the
resistivity and the
 Gilbert parameter $\alpha$ based on the scattering 
matrix approach applied to layered systems.\cite{LSYK11} 
The necessary average over many configurations 
of lattice displacements was taken by means of the
 super cell technique. In contrast to this the
 configurational average was determined using the 
Coherent Potential Approximation (CPA) within 
investigations using a Kubo-Greenwood-like linear 
expression for $\alpha$.\cite{EMKK11}
 The same approach to deal with the lattice
displacements was also used recently
within calculations of angle-resolved photo emission
 spectra (ARPES) on the basis of the 
one-step model of photo emission.\cite{BMM+13}

Another important contribution to the resistivity in the
case of magnetically ordered solids are thermally 
induced spin fluctuations.\cite{KDT+12} Again, the alloy 
analogy has been exploited extensively in the past
 when dealing with the impact of spin fluctuations
 on various response quantities. Representing a 
frozen spin configuration by means of super cell 
calculations has been applied for calculations of the
 Gilbert parameter for $\alpha$ \cite{LSYK11} as well as the 
resistivity or conductivity, respectively.\cite{LSYK11,GBK+12,KMWB14}~
Also, 
the CPA has been used for calculations of $\alpha$ \cite{MKE11} 
as well as the resistivity.\cite{AD93,KDT+12} A crucial point in this 
context is obviously the modeling of the temperature 
dependent spin configurations. Concerning this, rather 
simple models have been used,\cite{MKE11} but also quite 
sophisticated schemes. Here one should mention 
the transfer of data from Monte Carlo simulations 
based on exchange parameters calculated in an 
ab-initio way \cite{LKAG87} as well as work based on the 
disordered local moment (DLM) method.\cite{AD93,GPS+85}
Although, the standard DLM does not account for transversal 
spin components it nevertheless allows to represent 
the paramagnetic regime with no net magnetization in 
a rigorous way.
Also, for the magnetically ordered 
regime below the Curie-temperature it could be 
demonstrated that the uncompensated
DLM (uDLM) leads for many situations still to 
good agreement with experimental data on the 
so-called spin disorder contribution to the resistivity.\cite{KDT+12,AD93} 

In the following we present technical details and 
extensions of a scheme that was already used before
 when dealing with the temperature dependence of 
response quantities on the basis of Kubo's response 
formalism. Various applications will be presented 
for the conductivity and Gilbert 
damping parameter accounting simultaneously for various types of disorder.

\section{Theoretical framework}

\subsection{Configurational average for linear response functions}

Many important quantities in spintronics can be
 formulated by making use of linear
 response formalism. Important examples for this are 
the electrical conductivity,\cite{Vel69,But85}
 the spin conductivity \cite{LGK+11} or the
 Gilbert damping parameter.\cite{BTB08,EMKK11} 
Restricting here for the sake of brevity to the symmetric part of the 
corresponding response tensor $\chi_{\mu\nu}$ 
this can be expressed by a correlation function of the form:
%
\begin{equation}
\label{eq:sig}
\chi_{\mu\nu}  
\propto \, {\rm{Tr}}\,
\big\langle \hat{A}_{\mu} \, \Im G^+ \, \hat{A}_{\nu} \, \Im G^+ 
\big\rangle_{\rm c} \;.
\end{equation}
%
It should be stressed that this not a real restriction as the
scheme described below has been used successfully when
dealing with the impact of finite temperatures 
 on the anomalous Hall conductivity of Ni.\cite{KCME13}
In this case  the more complex Kubo-St\v{r}eda- or 
Kubo-Bastin formulation for the 
full response tensor has to be used.\cite{CB01} 

The vector operator $\hat{A}_{\mu}$  in  Eq.~(\ref{eq:sig}) stands
for example in case of the electrical conductivity $\sigma_{\mu\nu}$  for the
current density operator   $\hat{j}_{\mu}$  \cite{But85} 
while in case of the Gilbert damping parameter
$\alpha_{\mu\nu}$ it stands for the torque operator 
 $\hat{T}_{\mu}$.\cite{SF05,EMKK11}
Within the Kubo-Greenwood-like equation 
(\ref{eq:sig}) the electronic structure of the investigated
system is represented
 in terms of its retarded Green function 
$ G^+ ( \vec r,\vec r\,',E) $.
 Within  multiple scattering theory
or the KKR (Korringa-Kohn-Rostoker) formalism, 
$G^+(\vec{r},\vec r\,',E)$ can be written as:\cite{FS80,Wei90,ebe00}
%
\begin{eqnarray}
\label{eq:GFUN}
 G^+(\vec{r},\vec r\,',E)& = &
\sum_{\Lambda \Lambda'}
Z^{m}_{\Lambda}(\vec r,E)
 \tau^{mn}_{\Lambda \Lambda'}(E)
Z^{n\times}_{\Lambda'}(\vec r\,',E)\\
&&-
 \delta_{mn} \sum_{\Lambda }
   Z^{n}_{\Lambda}(\vec r,E) J^{n\times}_{\Lambda'}(\vec r'_{} ,E) \Theta(r_n'-r_n) 
 \nonumber \\
&& \hspace{8.3ex}
+  J^{n}_{\Lambda}(\vec r,E) Z^{n\times}_{\Lambda'}(\vec r' ,E) \Theta(r_n-r_n')
\;.
 \nonumber
\end{eqnarray}
%
Here  $\vec{r},\vec{r}'$ refer  to points  within atomic  volumes around
sites $\vec R_m,  \vec R_n$, respectively, with $Z^{n}_{\Lambda}(\vec
r,E)=Z_{\Lambda}(\vec r_n,E) = Z_{\Lambda}(\vec r-\vec R_n ,E) $ being a
function centered at site $\vec R_n$.
Adopting a fully relativistic 
formulation\cite{Wei90,ebe00} for Eq.~(\ref{eq:GFUN})
one gets   in a natural way  access to all 
spin-orbit induced properties as for example 
the anomalous and spin Hall conductivity\cite{CB01,LKE10b,LGK+11}
or Gilbert damping parameter.\cite{EMKK11}
In this case,
 the functions $Z^{n}_{\Lambda}$ and $J^{n}_{\Lambda}$ stand for the
regular and irregular, respectively,
solutions to the single-site Dirac equation
for site $n$ with the associated 
single-site scattering t-matrix $ t^n_{\Lambda \Lambda'}$.
The corresponding scattering path operator 
$\tau^{nn'}_{\Lambda \Lambda'}$
accounts for all scattering events connecting
 the sites $n$ and $n'$. Using a suitable spinor
 representation for the basis functions the combined quantum number
 $\Lambda=(\kappa,\mu)$  stands for the
 relativistic spin-orbit and magnetic quantum numbers $\kappa$ and $ \mu$, 
respectively.\cite{Ros61,Wei90,ebe00}

As was demonstrated by various authors \cite{Vel69,But85,TKD+02b}
representing the electronic structure 
 in terms of the Green function $ G^+ ( \vec r,\vec r\,',E) $
allows to account
for chemical disorder in a random 
alloy by making use of a suitable alloy theory.
In this case  $\langle ... \rangle_{\rm c}$
 stands for the configurational average for a substitutional alloy
concerning the site occupation.
Corresponding expressions for the conductivity tensor
have been worked out by
 Velick\'y \cite{Vel69} and  Butler \cite{But85} using
 the single-site Coherent Potential Approximation
(CPA) that include in particular the so-called vertex corrections.

The CPA can be used to deal with chemical 
but also with any other type of disorder.
In fact, making use of the different  time scales connected 
with the electronic propagation and spin fluctuations the alloy analogy
is exploited when dealing with finite temperature magnetism
on the basis of the disordered local moment (DLM) model.\cite{GPS+85,SG92b}
Obviously,  the same approach can be used when dealing with 
response tensors at  finite temperatures. In connection
with the conductivity 
this is often called adiabatic approximation.\cite{JM90}
Following this philosophy, 
the CPA has been used recently also when calculating response tensors 
using Eq.~(\ref{eq:sig}) with disorder in the system caused by
thermal lattice vibrations \cite{EMKK11,KCME13} 
as well as spin fluctuations.\cite{KDT+12,MKWE13}

\subsection{Treatment of thermal lattice displacement}
\label{sec-thermal-lattice-displacements}
\begin{figure}[hbt] 
   \includegraphics[angle=0,width=0.9\linewidth,clip]{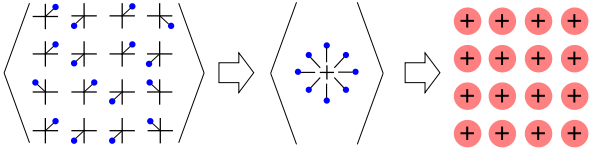}
%
\caption{\label{CPA-displacement} 
Configurational averaging for thermal lattice displacements: 
the continuous distribution 
$P(\Delta \vec{R}_n(T))$  for the atomic displacement 
vectors is replaced by a discrete set of vectors 
$\Delta \vec{R}_v(T)$ occurring 
with the probability $x_v$. The configurational average for this
 discrete set of displacements is made using the CPA leading 
to a periodic effective medium.}
\end{figure}

A way to account for the  impact of the thermal displacement of atoms
from their equilibrium positions,
i.e.\ for thermal lattice vibrations, 
on the electronic structure is to set up
a representative  displacement configuration for the
atoms within an enlarged unit cell (super-cell  technique). 
In this case one has to use either a
very large super-cell or to take the average
over a set of super-cells.
Alternatively, one may make use of the alloy analogy
for the averaging problem. 
This allows in particular to restrict 
to the standard unit cell.
Neglecting the correlation
 between the thermal displacements of neighboring
atoms from their equilibrium positions 
the properties of the thermal averaged system
can be deduced by making use of the single-site
CPA. This basic idea is illustrated by 
Fig.\ \ref{CPA-displacement}.
%
%
To make use of this scheme a discrete 
set of $N_{v}$ displacement vectors 
$\Delta \vec{R}^q_v(T)$ 
with probability $x^q_v$
($v=1,..,N_{v}$)
is constructed   for each basis atom $q$
within the standard unit cell that is conform with the local symmetry and
 the temperature dependent root mean  square 
displacement $(\langle u^2\rangle_T)^{1/2}$ according to:
%
\begin{equation}
\label{eq:displacement}
\frac{1}{N_{v}} 
\sum_{v=1}^{N_{v}} | \Delta \vec{R}^q_v(T) |^2 = \langle u_q^2\rangle_T \;.
\end{equation}
%
In the general case, the mean square displacement along the direction
  $\mu$ ($\mu = {x,y,z}$) of the atom $i$ can be either taken from
experimental data or represented by the expression based on the
phonon calculations \cite{Boe83}  

\begin{equation}
\label{eq:MSD}
\langle
 u^2_{i,\mu}
\rangle_T = \frac{3\hbar}{2M_i}
\int_0^{\infty} d\omega
g_{i,\mu}(\omega)\frac{1}{\omega}
{\rm{coth}}\frac{\hbar\omega}{2k_{\rm B}T}  \;,
\end{equation}
%
where  
$h=2\pi\hbar$ the Planck constant,  $k_{\rm B}$ the Boltzmann constant,
$g_{i,\mu}(\omega)$ is a partial phonon density of states. \cite{Boe83}
On the other hand, a rather good estimate for the root mean square
displacement can be obtained using  Debye's theory. In this
case, for systems with one atom per unit cell, Eq.\ (\ref{eq:MSD}) can
be reduced to the expression: 
%
%
%
\begin{equation}
\label{eq:Debye}
\langle u^2\rangle_T =
\frac{1}{4}\frac{3h^2}{\pi^2Mk_{\rm B}
\Theta_D}\left[\frac{\Phi(\Theta_D/T)}{\Theta_D/T}
+ \frac{1}{4}\right]
\end{equation}
%
with $\Phi(\Theta_D/T)$ the Debye function and
$\Theta_D$ the Debye temperature \cite{GMMP83}.
Ignoring the zero temperature term $1/4$ and
assuming a frozen potential for the atoms, the situation can be dealt
with in full analogy to the treatment of disordered alloys on the basis
of the CPA. The probability $x_v$ for a specific displacement
$v$ may normally be chosen as $1/N_{v}$. 
The Debye temperature $\Theta_D$ used in Eq.~(\ref{eq:Debye}) can be
either taken from  experimental data or calculated
representing it in terms of the elastic constants \cite{FBS01}. In general 
the latter approach 
should give more reliable results in the case of multicomponent systems.

To simplify notation we restrict in the following to
systems with one atom per unit cell. 
The index $q$ numbering sites in the unit cell can therefore
be dropped, while the  index $n$ numbers the lattice sites.

Assuming a rigid displacement of the atomic potential 
in the spirit of the rigid muffin-tin approximation \cite{PZDS97,Lod76}
the corresponding single-site t-matrix $ \underline{t}^{\rm loc}$
with respect to the local frame of reference
connected with the displaced atomic position
is unchanged. 
With respect to the global frame of  reference
connected with the equilibrium atomic positions $ \vec{R}_n$, however,
the corresponding t-matrix $ \underline{t}$
is given by the transformation:
%
\begin{equation}
\label{eq:U-trans}
\underline{t} = \underline{U}(\Delta \vec{R})\,\underline{t}^{\rm loc}\,
                     \underline{U}(\Delta \vec{R})^{-1} \;.
\end{equation}
%
The so-called U-transformation matrix $\underline{U}(\vec{s})$ 
is given  in its
non-relativistic form 
by:\cite{Lod76,PZDS97}
%
\begin{equation}
\label{eq:U-trans-matrix}
U_{LL'}(\vec{s}) = 
4\pi \sum_{L''}i^{l+l''-l'}\, C_{LL'L''}\, j_{l''}(|\vec{s}|k)\, Y_{L''}(\hat{s})
\;.
\end{equation}
%
Here $L=(l,m)$ represents the non-relativistic 
angular momentum quantum numbers,
$j_{l}(x)$ is a spherical Bessel function,
$Y_{L}(\hat{r})$ a real spherical harmonics, 
$C_{LL'L''}$ a corresponding  Gaunt number
and $k=\sqrt{E}$ is the electronic wave vector.
The relativistic version of the U-matrix 
is obtained by a standard Clebsch-Gordan transformation.\cite{Ros61}

The various displacement vectors $\Delta \vec{R}_v(T)$ 
can be used to determine the properties of a
 pseudo-component of a pseudo alloy.
Each of the
 $N_{v}$ pseudo-components with $|\Delta \vec{R}_v(T)| = \langle
 u^2\rangle_T ^{1/2}$ is characterized
by a corresponding  U-matrix 
 $\underline{U}_v$ and
t-matrix  $\underline{t}_v$. As for a substitutional alloy
the configurational
average can be determined by solving the
multi-component CPA equations 
within the global frame of  reference:
%
\begin{eqnarray}
\label{eq:CPA1}
\underline{\tau}_{{\rm CPA}}^{nn} &= &
\sum_{v=1}^{N_{v}}
x_v \underline{\tau}_{v}^{nn} 
\\
%
\label{eq:CPA2}
\underline{\tau}_{v}^{nn}& = &
\big[ 
    (\underline{t}_{v})^{-1}
-   (\underline{t}_{{\rm CPA}})^{-1}
+   (\underline{\tau}_{{\rm CPA}}^{nn})^{-1}
\big]^{-1}
\\
\label{eq:CPA3}
\underline{\tau}_{{\rm CPA}}^{nn}
 & = & \frac{1}{\Omega_{{\rm BZ}}}
 \int_{\Omega_{\rm BZ} } d^{3}k 
\left[ (\underline{t}_{{\rm CPA}})^{-1} 
      - \underline{G}(\vec{k},E)  \right]^{-1}  \; ,
\end{eqnarray}
%
where the underline indicates 
matrices with respect to the
combined index $ \Lambda$.
As it was pointed out in the previous work \cite{MKWE13},
the cutoff for the angular momentum expansion in these
calculations should be taken $l \geq l_{max} + 1$ with the $l_{max}$
value used in the calculations for the non-distorted lattice.

The first of these CPA equations represents the
 requirement for the mean-field CPA medium
that  embedding of a component $v$ 
should lead in the average to no additional scattering.
Eq.~(\ref{eq:CPA2}) gives the scattering path operator for the
embedding of the component $v$ into the
CPA medium while Eq.~(\ref{eq:CPA3}) 
gives the  CPA scattering path operator 
in terms of a Brillouin
zone integral with $ \underline{G}(\vec{k},E)$ the so-called 
KKR structure constants.

Having solved the CPA equations
the linear response quantity of interest
may be calculated using Eq.~(\ref{eq:sig}) as for an
ordinary substitutional alloy.\cite{Vel69,But85}
This implies that one also have to 
deal with the so-called vertex corrections\cite{Vel69,But85}
that take into account that one has to
deal with a configuration average of the type
$\langle \hat{A}_{\mu} \, \Im G^+ \, 
         \hat{A}_{\nu} \, \Im G^+ 
 \rangle_{\rm c} $
 that in general will differ from the
simpler product  
$\langle \hat{A}_{\mu} \, \Im G^+ \, 
 \rangle_{\rm c} 
 \langle
          \hat{A}_{\nu} \, \Im G^+ 
 \rangle_{\rm c} $.

\subsection{Treatment of thermal spin fluctuations}

\begin{figure}[hbt]
 \begin{center}
   \includegraphics[angle=0,width=0.9\linewidth,clip]{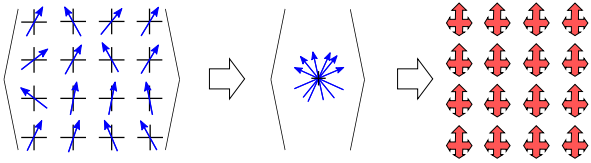}
 \end{center}
  \caption{
\label{CPA-fluctuation}  
Configurational averaging for thermal spin fluctuations: 
the continuous distribution $P(\hat{e}_n)$ for the orientation of
 the magnetic moments is replaced by a discrete set of 
orientation vectors $\hat{e}_f$ occurring with a probability $x_f$.
 The configurational average for this discrete set of 
orientations is made using the CPA leading to a periodic effective medium.
}
\end{figure}

As for the disorder connected with
thermal displacements the impact of disorder
due to thermal spin fluctuations
may be accounted for by use of the super-cell
technique.~
Alternatively one may again use the alloy analogy
and determine the necessary configurational average
by means of the CPA 
as indicated 
in Fig.\ \ref{CPA-fluctuation}.
%
%
As for the thermal displacements
in a first step a set of representative 
orientation vectors $\hat{e}_f$ 
(with $f=1,...,N_f$) for the local magnetic
moment is introduced (see below).
Using the rigid spin approximation 
the spin-dependent
part  $B_{\rm xc}$ of the exchange-correlation potential
does not change for the local frame of  reference fixed to
the magnetic moment 
when the moment is oriented along an 
orientation vector  $\hat{e}_f$ .
This implies that the single-site t-matrix
 $ \underline{t}_f^{\rm loc}$ in the local frame is the same
for all orientation vectors. With respect 
to the common global frame 
that is used to deal with the multiple scattering
(see Eq.~(\ref{eq:CPA3}))
the t-matrix for a given orientation vector
is determined by:
%
\begin{equation}
\label{eq:R-trans}
\underline{t} = \underline{R}(\hat{e})\,\underline{t}^{\rm loc}\,
                \underline{R}(\hat{e})^{-1} \;.
\end{equation}
%
Here the transformation from the
local to the global frame of reference
is expressed by the rotation matrices $ \underline{R}(\hat{e})$
that are determined by the vectors $\hat{e}$
or corresponding Euler angles.\cite{Ros61}

Again the configurational average
for the pseudo-alloy can be obtained
by setting  up and solving 
CPA equations in analogy to
Eqs.~(\ref{eq:CPA1}) to (\ref{eq:CPA3}).

\subsection{Models of spin disorder}
\label{seq:Model-spin-disorder}

\begin{figure}[hbt]
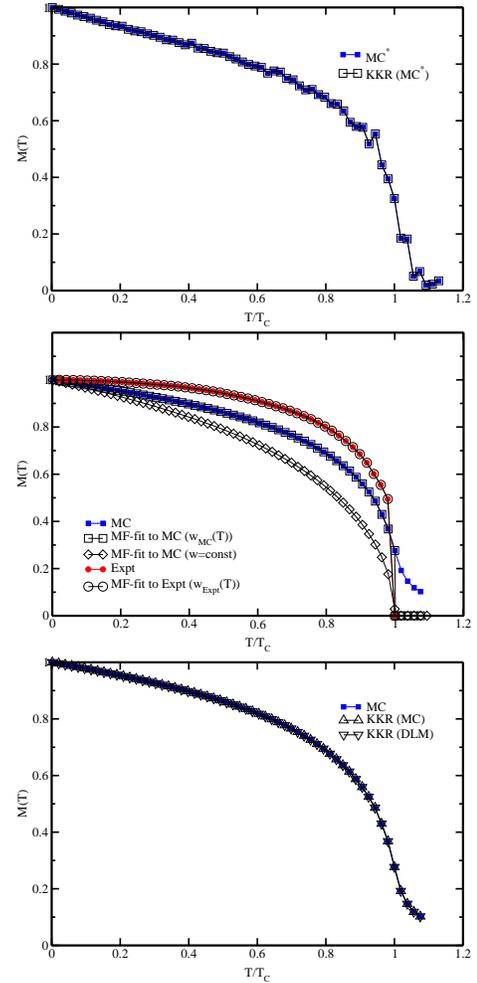

 \begin{center}
   \includegraphics[angle=0,width=0.7\linewidth,clip]{Fig_3a_Fe_M_T_atoms4096_5a.eps}
   \includegraphics[angle=0,width=0.7\linewidth,clip]{Fig_3b_Fe_M_T_fit_to_MC_mod.eps}
   \includegraphics[angle=0,width=0.7\linewidth,clip]{Fig_3c_Fe_M_T_smooth_5b.eps}
  \caption{\label{FIG-MAG-T}
Averaged reduced magnetic
moment 
$M(T) = \langle m_z \rangle_T / |\langle \vec{m} \rangle_{T=0}|$ 
along the z-axis as a function of the 
temperature $T$.
Top: results of Monte Carlo simulations using
scheme MC* (full squares) compared with results of
subsequent KKR-calculations (open squares).
Middle: results of Monte Carlo simulations using
scheme MC (full squares) compared with results
using a mean-field fit with a constant Weiss field
$w_{\rm MC}(T_C)$ (open diamonds) and a temperature dependent
Weiss field $w_{\rm MC}(T)$ (open squares).
In addition experimental data (full circles) together
with a corresponding  mean-field fit obtained for a 
temperature dependent
Weiss field $w_{\rm exp}(T)$. 
Bottom: results of Monte Carlo simulations using
scheme MC (full squares) compared with results
subsequent KKR-calculations 
using the MC (triangles up)
and a corresponding DLM (triangle down) spin configuration,
respectively.
%
}
 \end{center}
\end{figure}

The central problem with the scheme described above is obviously
 to construct a realistic and
 representative set of orientation vectors $\hat{e}_f$
and probabilities  ${x}_f$
 for each temperature $T$. A rather appealing approach is to calculate 
the exchange-coupling parameters $J_{ij}$ of a system
 in an ab-initio way \cite{LKAG87,USPW03,EM09a} 
and to use them in subsequent Monte Carlo simulations. 
Fig.\ \ref{FIG-MAG-T} (top) shows results for 
the temperature dependent  average reduced
magnetic moment  
of  corresponding simulations for bcc-Fe
 obtained for a periodic cell with 4096 atom sites. 
%
%
The full line gives the 
value for the reduced magnetic moment 
$M_{\rm MC^*}(T) = \langle m_z \rangle_T /m_0$ 
projected on the z-axis for the last Monte Carlo step
($\hat z$ is the orientation of the total moment, i.e.\
$\langle \vec{m}\rangle_T \| \hat z $; the saturated magnetic moment at
$T = 0$~K is $m_0 = |\langle \vec{m} \rangle_{T=0}|$). This scheme is
 called ${\rm MC^*}$ in the following. 
In spite of the rather large number of sites (4096) the curve
 is rather noisy in particular when 
approaching the Curie temperature. 
Nevertheless, the spin configuration of 
the last MC step was used as an input for subsequent 
SPR-KKR-CPA calculations using
 the orientation vectors $\hat{e}_f$ with the probability $x_f=1/N_f$
 with $N_f = 4096$. 
As Fig.\ \ref{FIG-MAG-T} (top) shows, the 
temperature dependent reduced magnetic moment $M_{\rm KKR(MC^*)}(T)$ deduced 
from the electronic structure calculations follows one-to-one the
 Monte Carlo data $M_{\rm MC^*}(T)$. This is a very encouraging result for 
further applications (see below) as it demonstrates that the 
CPA although being a mean-field method and used here in its single-site 
formulation is nevertheless capable to reproduce results of MC simulations
 that go well beyond  the mean-field level.

However, using the set of vectors $\hat{e}_f$ of scheme 
MC* also for calculations of the 
Gilbert damping parameters $\alpha$ as a 
function of temperature led to extremely
 noisy and unreliable curves for $\alpha(T)$. For that reason an average has 
been taken over many MC steps 
(scheme MC) leading to a much smoother
 curve for $M_{\rm MC}(T)$ as can be seen from 
Fig.\ \ref{FIG-MAG-T} (middle) with
 a Curie temperature $T_{\rm C}^{\rm MC}=1082$~K.
 As this enlarged set of vectors 
$\hat{e}_f$ got too large to be used directly  in subsequent 
SPR-KKR-CPA calculations, a scheme was worked out
 to get a set of vectors $\hat{e}_f$ and 
probabilities $x_f$ that is not too large 
but nevertheless leads to smooth curves 
for $M(T)$. 

The first attempt was to use
 the Curie temperature $T_{\rm C}^{\rm MC}$ to 
deduce a corresponding temperature independent Weiss-field $w(T_{\rm C})$ 
on the basis of the standard
 mean-field relation:
%
\begin{equation}
\label{eq:wTC}
w(T_{\rm C}) = \frac{3k_{\rm B}T_C}{m_0^2} \;.
\end{equation}
%
 This leads to a reduced magnetic moment 
curve $M_{\rm MF}(T)$ that shows by construction the same Curie temperature as
 the MC simulations. 
For temperatures between $T=0$~K  and $T_{\rm C}$, however, the
 mean-field reduced magnetic moment $M_{\rm MF}(T)$ is well below the MC curve
(see Fig.\ \ref{FIG-MAG-T} (middle) ).

As an 
alternative to this simple approach we introduced a temperature dependent
 Weiss field $w(T)$.
This allows to describe the temperature dependent magnetic properties 
using the results obtained beyond the mean-field approximation. 
At the same time the calculation of the statistical average can be performed
treating the model Hamiltonian in terms of the mean field theory. 
For this reason the reduced magnetic moment  $M(T)$, being a solution
of equation (see e.g.\ \cite{Tik64})  
%
\begin{equation}
\label{eq:xf}
 M(T)  = L\left(\frac{w m_0^2 M(T)}{k_{\rm B}T}\right) \;,
\end{equation}
%
was fitted to that obtained  from MC simulations $M_{\rm MC}(T)$
with the Weiss field $w(T)$ as a fitting parameter, such that 
\begin{equation}
\label{eq:xf-a}
\lim_{w \to w(T)} M(T)  = M_{MC}(T) \;,
\end{equation}
with $L(x)$ the Langevin function.

\begin{figure}[hbt]
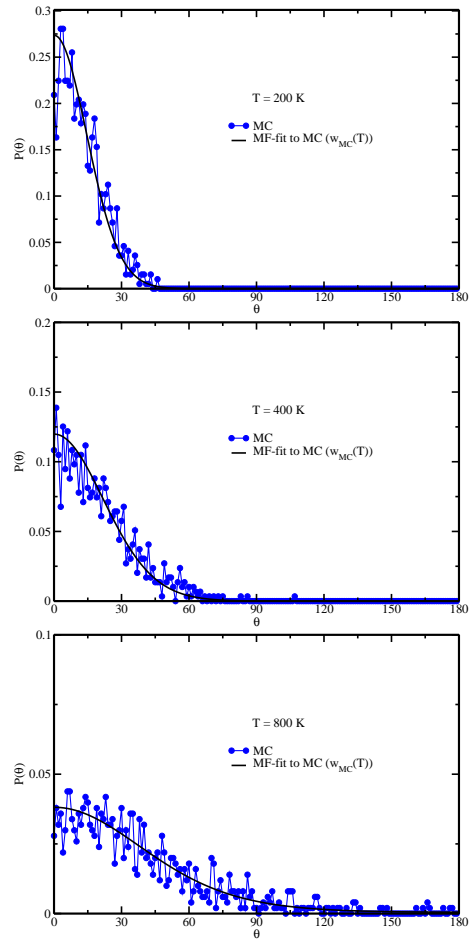

 \begin{center}
   \includegraphics[angle=0,width=0.7\linewidth,clip]{Fig_4a_Fe_M_THETA_200K.eps}
   \includegraphics[angle=0,width=0.7\linewidth,clip]{Fig_4b_Fe_M_THETA_400K.eps}
   \includegraphics[angle=0,width=0.7\linewidth,clip]{Fig_4c_Fe_M_THETA_800K.eps}
 \caption{\label{FIG-MAG-PROB} 
Angular distribution $P(\theta) $ 
of the atomic magnetic moment $\vec{m}$
obtained from Monte Carlo simulations (MC)
for the temperature $T= 200 $, $400 $, and $800$~K
compared with field mean-field (MF) data, $x_f$, (full line) 
obtained by fitting using a
temperature dependent Weiss field $w(T)$ (Eq.\ \ref{eq:xf}). 
}
 \end{center}
\end{figure}

The corresponding temperature dependent probability
 $x(\hat{e})$ for an atomic magnetic moment to be oriented along $\hat{e}$ is 
proportional to $\exp(-w(T) \hat z \cdot \hat e/k_{\rm B}T)$ (see,
e.g. \cite{Tik64}). 
 To calculate this value we 
used $N_{\theta}$ and $N_{\phi}$ points for
 a regular grid for the spherical angles $\theta$ and $\phi$ corresponding 
to the vector $\hat{e}_f$:
\begin{equation}
\label{eq:xf-b}
       x_f = \frac{\exp(-w(T) \hat z \cdot \hat{e}_{f}/k_{\rm B}T)}
  {\sum_{f'} \exp(-w(T) \hat z \cdot \hat{e}_{f'}/k_{\rm B}T)}   \;.
\end{equation}
 Fig.\ \ref{FIG-MAG-PROB} shows for three different 
temperatures the $\theta$-dependent behavior of $x(\hat{e})$.
%
%
As one notes, the MF-fit to the MC-results perfectly reproduces
 these data for all temperatures.
 This applies of course not only for the angular resolved 
 distribution of the magnetic moments shown in 
Fig.\ \ref{FIG-MAG-PROB} but also for the average 
reduced magnetic moment recalculated using Eq.(\ref{eq:xf}), shown in
Fig.\ \ref{FIG-MAG-T}. Obviously, the MF-curve $M_{\rm MF(MC)}(T)$ 
obtained using the temperature dependent Weiss field parameter $w(T)$ perfectly reproduces the
 original $M_{\rm MC}(T)$ curve. The great advantage 
of this fitting procedure is that it 
allows to replace the MC data set with a large
 number $N_f^{\rm MC}$ of orientation vectors 
$\hat{e}_f$ (pointing in principle into any direction)
 with equal probability $x_f=1/N_f^{\rm MC}$ 
by a much smaller data set with 
$N_f= N_\theta \,  N_\phi$ 
with $x_f$
 given by Eq.~(\ref{eq:xf-b}).

%
%
Accordingly, the reduced
 data set can straight forwardly
 be used for subsequent electronic structure calculations.
Fig.\ \ref{FIG-MAG-T} (bottom) shows that the calculated 
temperature dependent
 reduced magnetic moment $M_{\rm KKR-MF(MC)}(T)$ agrees perfectly
 with the reduced magnetic moment 
$M_{\rm MC}(T)$ given by the underlying MC simulations.

The DLM method has the appealing feature that it combines
 ab-initio calculations
 and thermodynamics in a coherent way.
 Using a non-relativistic formulation, it 
was shown that the corresponding averaging 
over all orientations of the individual 
atomic reduced magnetic moments can be mapped onto a
 binary pseudo-alloy with one 
pseudo-component having up- and 
 downward orientation of the spin moment  with concentrations 
$x_{\uparrow}$ and $x_{\downarrow}$, respectively.\cite{AD93,Aka98}
 For a fully relativistic formulation, 
with spin-orbit coupling included, this 
simplification cannot be justified anymore
 and a proper average has to be taken over
 all orientations.\cite{SOR+04} 
As we do not perform DLM calculations but use 
here only the DLM picture to 
represent MC data, this complication is 
ignored in the following. Having the set of 
orientation vectors $\hat{e}_f$ determined by
 MC simulations the corresponding 
concentrations $x_{\uparrow}$ and $x_{\downarrow}$
 can straight forwardly be fixed for each 
temperature by the requirement:
%
\begin{equation}
\label{eq:DLM-requ}
\frac{1}{N_f} \sum_{f=1}^{N_f} 
\hat{e}_f = 
x_{\uparrow} \hat z +  x_{\downarrow}(- \hat z )
 \;,
\end{equation}
%
with $x_{\uparrow} + x_{\downarrow} = 1$.
Using this simple scheme electronic structure calculations have been performed 
for a binary alloy having collinear magnetization. The resulting 
reduced magnetic moment $M_{\rm KKR-DLM(MC)}(T)$ is shown in 
Fig.\ \ref{FIG-MAG-T} (bottom). As one notes, again the original MC 
results are perfectly reproduced. This implies that when calculating 
the projected reduced magnetic moment $M_{\rm z}$ that is determined by
the averaged Green function $\langle G \rangle$  the 
transversal magnetization has hardly any impact.

Fig.\ \ref{FIG-MAG-T} (middle) gives also 
experimental data for the $M(T)$.\cite{CG71}
While the experimental Curie-temperature 
 $T_{\rm C}^{\rm exp}=1044$~K\cite{CG71} 
is rather well reproduced by the 
MC simulations $T_{\rm C}^{\rm MC}=1082$~K 
one notes that the MC-curve $M_{\rm MC}(T)$ 
is well below the experimental curve. In particular, 
$M_{\rm MC}(T)$ drops too 
fast with increasing $T$ in the low temperature regime and does not show the 
$T^{3/2}$-behavior. 
The reason for this is that the MC simulations do not 
properly account for the low-energy long-ranged spin wave excitations
responsible for the low-temperature magnetization variation.
 Performing ab-initio calculations for the spin wave energies and using
 these data for the calculation of $M(T)$ much better agreement with 
experiment can indeed be obtained in the low-temperature regime than with
 MC simulations.\cite{HEPO98} 

As the fitting scheme sketched above needs only 
the temperature reduced magnetic moment $M(T)$ as input it can be applied not only 
to MC data but also to experimental data. 
Fig.\ \ref{FIG-MAG-T} shows that the mean field fit $M_{\rm MF(exp)}(T)$ again 
perfectly fits the experimental
 reduced magnetic moment curve $M_{\rm exp}(T)$. Based on 
this good agreement this 
corresponding data set $\{\hat{e}_f,x_f\}$ 
has also been used 
for the calculation of response tensors (see below).

An additional much 
simpler scheme to simulate the experimental $M_{\rm exp}(T)$ curve is to assume the 
individual atomic moments to be distributed on a cone, i.e.\  with $N_\theta=1$
 and $N_{\phi}>>1$.\cite{MKE11}
 In this case the opening angle $\theta(T)$ of the cone is chosen 
such to reproduce $M(T)$. In contrast to the standard DLM picture, 
this simple scheme allows already to account for transversal components
 of the magnetization. Corresponding results for response tensor calculations will be shown below.

Finally, it should be stressed here that the various spin configuration models
discussed above assume a rigid spin moment, i.e.\ its magnitude does not change with
temperature nor with orientation. In contrast to this 
Ruban et al.\ \cite{RKMJ07}
use a longitudinal spin fluctuation Hamiltonian with the corresponding parameters 
derived from ab-initio calculations.
As a consequence, subsequent Monte Carlo simulations based on this Hamiltonian
account in particular for longitudinal fluctuations of the spin moments.
A similar approach has been used by Drchal et al. \cite{DKT13,DKT13a}
leading to good agreement with the results of Ruban et al.
However, the scheme used in these calculations does not supply in a
straightforward manner the necessary input for temperature dependent
transport calculations.
This is different from the work of Staunton et al. \cite{SBS+14} who
performed self-consistent relativistic DLM calculations without the
restriction to a collinear spin configuration. This approach in
particular accounts in a self-consistent way for longitudinal spin
fluctuations.

\subsection{Combined chemical and thermally induced disorder}

The various types of disorder discussed above may be combined 
with each other as well as with chemical i.e.\  substitution 
disorder. In the most general case a pseudo-component $(vft)$ 
is characterized by its chemical  atomic type $t$, the spin 
fluctuation $f$ and lattice displacement $v$. Using the rigid 
muffin-tin and rigid spin approximations, the single-site 
t-matrix $t^{\rm loc}_t $ in the local frame is independent from the 
orientation vector $\hat{e}_f$ and displacement vector 
$\Delta \vec{R}_{v}$, 
and coincides with $ \underline{t}_t$ for the atomic type $t$. With 
respect to the common global frame one has accordingly
 the t-matrix:
%
\begin{equation}
\label{eq:tvft}
 \underline{t}_{vft} =     \underline{U}(\Delta \vec{R}_{v})\,
                     \underline{R}(\hat{e}_{f})\,
                     \underline{t}_t \,
                     \underline{R}(\hat{e}_{f})^{-1}
                     \underline{U}(\Delta \vec{R}_{v})^{-1} 
\;.
\end{equation}
%
With this the corresponding CPA equations are identical
 to 
Eqs.~(\ref{eq:CPA1}) to (\ref{eq:CPA3})
 with the index $v$ replaced by the 
combined index $(vft)$. The corresponding
 pseudo-concentration $x_{vft}$ combines the concentration
 $x_t$ of the atomic type $t$ with the probability for 
the orientation vector $\hat{e}_f$ and displacement vector $\Delta \vec{R}_{v}$.

\section{Computational details}

The electronic structure of the investigated ferromagnets 
bcc-Fe   and fcc-Ni, 
has been calculated self-consistently using the 
spin-polarized  relativistic 
KKR (SPR-KKR)
band structure 
 method.\cite{EKM11,SPR-KKR6.3} 
For the exchange correlation potential the parametrization 
as given by Vosko et al.\ \cite{VWN80} has been used. 
The angular-momentum cutoff of $l_{max} = 3$  was used in the KKR multiple
scattering expansion.
The lattice parameters
 have been set to the experimental values. 

In a second step the
 exchange-coupling parameters $J_{ij}$ have been calculated
 using the so-called Lichtenstein formula.\cite{LKAG87} Although the 
SCF-calculations have been done on a fully-relativistic level the
 anisotropy of the exchange coupling due to the spin-orbit coupling has
 been neglected here. Also, the small influence of the magneto-crystalline
 anisotropy for the subsequent Monte Carlo (MC) simulations has been 
ignored, i.e.\  these have been based on a classical Heisenberg Hamiltonian. 
The MC simulations were done in a standard way using the Metropolis algorithm
 and periodic boundary conditions.
 The theoretical Curie temperature $T_{\rm C}^{\rm MC}$ has been 
deduced from the maximum of the magnetic susceptibility.

The temperature dependent spin configuration obtained during a MC 
simulation has been used to construct 
a set of orientations $\hat{e}_f$ 
and probabilities  ${x}_f$ according to the  schemes 
MC* and MC described in section \ref{seq:Model-spin-disorder}
to 
be used within subsequent SPR-KKR-CPA calculations
 (see above). For the corresponding 
calculation of the reduced magnetic moment the potential obtained from the 
SCF-calculation for the perfect ferromagnetic state ($T=0$K) has been used.
The calculation for the electrical conductivity as well as the Gilbert
 damping parameter has been performed as 
described elsewhere.\cite{KLSE11,MKWE13}

\section{Results and discussion}

\subsection{Temperature dependent conductivity}

  Eq.~(\ref{eq:sig})  has been used together with the various 
schemes described above to calculate the temperature 
dependent longitudinal resistivity  $ \rho (T) $ 
 of the pure
ferromagnets Fe, Co and Ni. In this case obviously
disorder due to thermal displacements of  the atoms
as well as spin fluctuations contribute to the resistivity.

\begin{figure}[hbt]
 \begin{center} \includegraphics[angle=0,width=0.8\linewidth,clip]{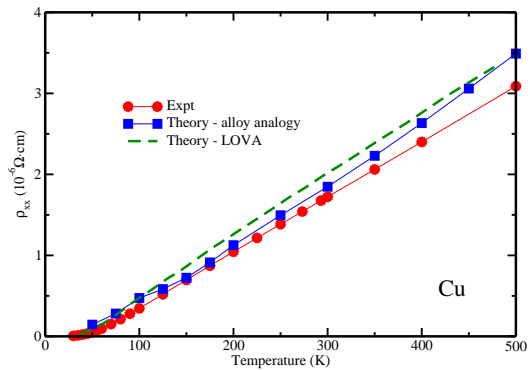}
  \caption{\label{FIG-RHO-Cu}
Temperature dependent longitudinal
 resistivity of fcc-Cu $ \rho(T)$
obtained by accounted for thermal vibrations 
as described in section \ref{sec-thermal-lattice-displacements} compared 
with corresponding experimental data.\cite{Bas82}
 In addition results are shown based on the LOVA (lowest order variational approximation) to the 
Boltzmann formalism.\cite{SS96}
}
 \end{center}
\end{figure}

To give an impression on the impact of the
thermal displacements alone
  Fig.\ \ref{FIG-RHO-Cu}
gives the temperature dependent 
 resistivity  $ \rho (T) $ 
of  pure Cu ($ \Theta_{\rm Debye}=315$ K)
that is found in very good agreement with
corresponding experimental data.\cite{Bas82}
%
%
This implies that the alloy analogy model
that ignores any inelastic scattering events
should in general lead to rather reliable results for the
resistivity induced by thermal displacements.
Accordingly, comparison with experiment should allow
for magnetically ordered systems
to find out the most appropriate model 
for spin fluctuations.

\begin{figure}[hbt]
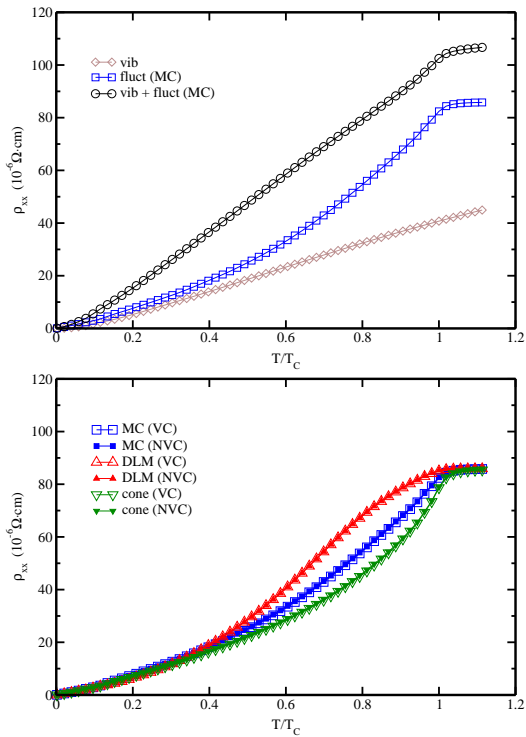

 \begin{center}
   \includegraphics[angle=0,width=0.8\linewidth,clip]{Fig_6a_Fe_rho_xx_vib_fluct_MC.eps}
   \includegraphics[angle=0,width=0.8\linewidth,clip]{Fig_6b_Fe_fluct_MC60_DLM_vs_all_dir.eps}
  \caption{\label{FIG-RHO-DISP-FLUCT}
Temperature dependent longitudinal
 resistivity of bcc-Fe $ \rho(T)$
obtained by accounted for thermal vibrations 
and spin fluctuations 
as described in section \ref{sec-thermal-lattice-displacements}.
Top:  accounting for 
vibrations                        (vib, diamonds),
spin fluctuations using scheme MC (fluct, squares) 
and both                          (vib+fluct, circles).
Bottom: accounting for spin fluctuations $\hat{e}_f = \hat{e}(\theta_f,\phi_f)$ 
using the schemes:  MC       (squares) with $0 \leq \theta_f \leq \pi; 0 \leq \phi_f \leq 2\pi$,
                    DLM(MC)  (triangles up) with $\theta_{f1} = 0, \theta_{f2} = \pi$,
and                 cone(MC) (triangles down) $\theta_f = \langle \theta_f
\rangle_T; 0 \leq \phi_f \leq 2\pi$. 
The full and open
symbols represent the results obtained
with the vertex corrections included (VC) 
 and excluded (NV), respectively.
}
 \end{center}
\end{figure}

 Fig.\ \ref{FIG-RHO-DISP-FLUCT} (top) 
shows theoretical results for $ \rho (T) $ of bcc-Fe
due to thermal displacements $ \rho_{v}(T) $,
spin fluctuations described by the scheme MC
$ \rho_{ \rm MC}(T)$ as
 well as the combination of the two
influences ($ \rho_{ \rm v,MC}(T)$).
%
%
First of all one notes that $ \rho_{ \rm v}(T) $
is not influenced within the adopted model
by the Curie temperature $T_{ \rm C}$
 but is determined
only by the Debye temperature.
$ \rho_{ \rm MC}(T) $, on the other hand,
reaches saturation for   $T_{ \rm C}$
as the spin disorder does not increase anymore
with increasing temperature in the
paramagnetic regime.
 Fig.\ \ref{FIG-RHO-DISP-FLUCT} also
shows that  $ \rho_{ \rm v}(T) $
and 
$ \rho_{ \rm MC}(T) $ 
are comparable for low temperatures
but $ \rho_{ \rm MC}(T) $ exceeds 
 $ \rho_{ \rm v}(T) $ more and more
for higher temperatures.
Most interestingly, however, 
the resistivity for the combined influence
of thermal displacements and spin fluctuations
$ \rho_{ \rm v,MC}(T) $ does 
not coincide  with the sum of $ \rho_{ \rm v}(T) $
and 
$ \rho_{ \rm MC}(T) $ 
but exceeds the sum for low temperatures
and lies below the sum when approaching
  $T_{ \rm C}$.

Fig.\ \ref{FIG-RHO-DISP-FLUCT} (bottom) 
shows the results   of three  different
calculations including the effect of
spin fluctuations 
as a function of the temperature.
The curve $ \rho_{ \rm MC}(T) $
is identical with that given in
Fig.\ \ref{FIG-RHO-DISP-FLUCT} (top)
based on Monte Carlo simulations.
The curves $ \rho_{ \rm DLM(MC)}(T)$
and 
$ \rho_{ \rm cone(MC)}(T)$
are based on a DLM- and cone-like representation
of the MC-results, respectively.
For all three cases results are given including
as well as ignoring the vertex corrections.
As one notes the vertex corrections
play a negligible role for all three
spin disorder models.
This is fully in line with the experience for
the longitudinal resistivity of disordered
transition metal alloys: as long as the
the states at the Fermi level have 
dominantly d-character the vertex corrections
can be neglected in general. On the other
hand, if the sp-character dominates inclusion of
vertex corrections may
alter the result in the order of 10 \%.\cite{BEWV94,TKDW04}

Comparing 
the DLM-result $ \rho_{ \rm DLM(MC)}(T)$ with
$ \rho_{MC}(T) $
one notes 
in contrast to the results for $ M(T) $
shown above (see Fig.\ \ref{FIG-MAG-T} (bottom))
quite an appreciable deviation.
This implies that the restricted
collinear representation
of the spin configuration implied by the DLM-model
introduces errors for the
configurational average that seem in general
to be unacceptable,
For the Curie temperature and beyond in the
paramagnetic regime 
$ \rho_{ \rm DLM(MC)}(T)$ and
$ \rho_{ \rm MC}(T) $
coincide, as it was shown formally before.\cite{KDT+12}

Comparing finally
$ \rho_{ \rm cone(MC)}(T)$ 
based on the conical representation of
the MC spin configuration with
$ \rho_{ \rm MC}(T) $
one notes that also this simplification
leads to quite strong deviations from the
more reliable result.
Nevertheless, one notes that 
$ \rho_{ \rm DLM(MC)}(T)$ 
agrees  with
$ \rho_{ \rm MC}(T) $
for the Curie temperature and also 
accounts to some
extent for the impact of the transversal 
components of the magnetization.

\begin{figure}[hbt]
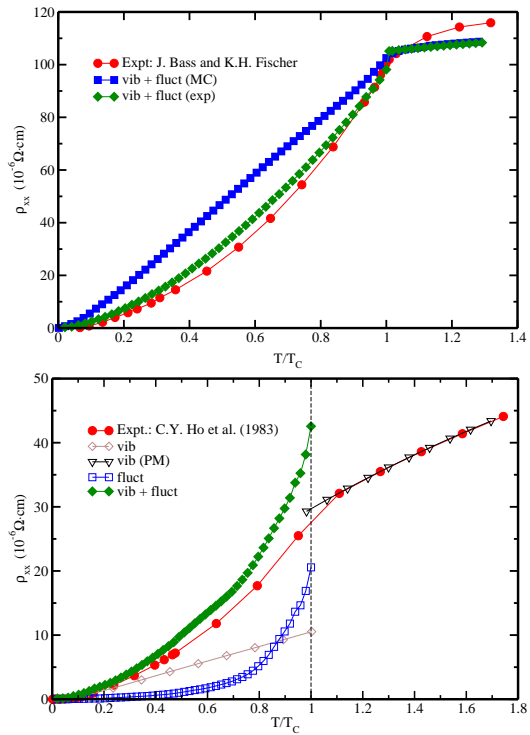

 \begin{center} 
  \includegraphics[angle=0,width=0.8\linewidth,clip]{Fig_7a_Fe_rho_xx_vib_fluct_MC_exp.eps}
  \includegraphics[angle=0,width=0.8\linewidth,clip]{Fig_7b_rho_xx_Ni_vib_fluct.eps}
 \end{center}
  \caption{\label{FIG-RHO-EXPT}
Top: Temperature dependent longitudinal
 resistivity of bcc-Fe $ \rho(T)$
obtained by accounted for thermal vibrations 
and spin fluctuations 
 using the scheme MC 
(vib+fluct(MC), squares)
and a mean-field fit to the experimental 
temperature
magnetic 
moment $M_{\rm exp} $
(vib+fluct(exp), diamonds)
compared with experimental data (circles).\cite{Bas82}
Bottom: corresponding results for fcc-Ni.
In addition results are shown accounting for thermal displacements
(vib) only for the ferromagnetic (FM) as well paramagnetic (PM) regime.
Experimental data have been taken from Ref.\ \onlinecite{HAW+83}.
}
\end{figure}

The theoretical results for bcc-Fe ($ \Theta_{\rm Debye}=420$ K) based
on the combined inclusion of the effects of
thermal displacements
and
spin fluctuations using the MC scheme
($ \rho_{ \rm v,MC}(T) $)
are compared in
Fig.\ \ref{FIG-RHO-EXPT} (top)
with experimental data ($ \rho_{ \rm exp}(T) $).
%
%
For the Curie temperature obviously a very good
agreement with experiment is found
while for lower temperatures $ \rho_{ \rm v,MC}(T) $
exceeds $ \rho_{ \rm exp}(T) $.
This behavior correlates well with
that of
the temperature dependent reduced magnetic moment $M(T)$
shown in  Fig.\ \ref{FIG-MAG-T} (middle).
The too rapid decrease of $M_{ \rm MC}(T) $
compared with experiment implies 
an essentially overestimated
spin disorder at any temperature leading in turn to a
too large resistivity 
$ \rho_{ \rm v,MC}(T) $.
On the other hand, using the 
temperature dependence of the
experimental reduced magnetic moment $M_{ \rm exp}(T) $
to set up the temperature dependent spin
configuration as described above
a very satisfying agreement is found with the
experimental resistivity data $ \rho_{ \rm exp}(T) $.
Note also that above $T_C$ the calculated resistivity
riches the saturation in contrast to the experimental data
where the continuing increase of $ \rho_{ \rm exp}(T) $ can be
attributed to the longitudinal spin fluctuations leading
to a temperature dependent distribution of local magnetic moments
on Fe atoms.\cite{RKMJ07}   
However, this contribution was not taken into account  
because of restriction in present calculations 
using fixed value for the local reduced magnetic moments.

Fig.\ \ref{FIG-RHO-EXPT} (bottom) shows corresponding results 
for the temperature dependent resistivity
of fcc-Ni ($ \Theta_{\rm Debye}=375$ K). For the ferromagnetic (FM) regime that the theoretical results 
are comparable in magnitude when only thermal displacements  ($ \rho_{ \rm v}(T) $)
or spin fluctuations ($ \rho_{ \rm MF}(T) $)
are accounted for.  
In the later case the mean field $w(T)$  has been fitted to the experimental
 $M(T)$-curve.    Taking both into account leads to 
a resistivity  ($ \rho_{ \rm v,MF}(T) $) 
 that are  well above the sum of the individual terms $ \rho_{ \rm v}(T) $ and
 $ \rho_{ \rm MF}(T) $.
Comparing $ \rho_{ \rm v,MF}(T) $ with experimental data 
 $ \rho_{ \rm exp}(T) $ our finding shows that the theoretical results
overshoots the experimental one the closer one comes to the critical temperature.
This is a clear indication that the assumption of a rigid spin moment
is quite questionable as the resulting contribution to the resistivity
due to spin fluctuations as much too small.
In fact the simulations of Ruban et al.\ \cite{RKMJ07} on the basis of a
longitudinal spin fluctuation Hamiltonian led on the case of fcc-Ni to a
strong diminishing of the average 
local magnetic moment when the critical temperature is approached from below
(about 20 \% compared to $ T=0$~K).
For bcc-Fe, the change is much smaller (about 3 \%) justifying on the case the
assumption of a rigid spin moment.
Taking the extreme point of view that the spin moment vanishes completely above
the critical temperature or the paramagnetic  (PM) regime only
thermal displacements have to be considered as a source for a finite resistivity.
Corresponding results are shown in
Fig.\ \ref{FIG-RHO-EXPT} (bottom) together with corresponding
experimental data. The very good agreement between both obviously suggests that
remaining spin fluctuations above the critical temperature are of minor importance
for the resistivity of fcc-Ni.


\subsection{Temperature dependent Gilbert damping parameter}

\begin{figure}[hbt]
 \begin{center}
   \includegraphics[angle=0,width=0.8\linewidth,clip]{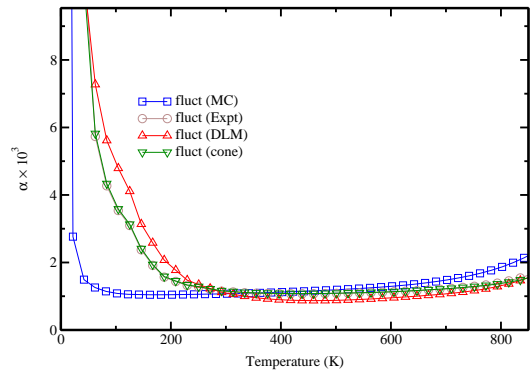}
  \caption{\label{FIG-GIL-THEORY} Temperature dependent Gilbert damping
    $\alpha(T)$ for bcc-Fe, obtained by accounted for thermal vibrations
    and spin fluctuations accounting for spin fluctuations using 
scheme MC (squares),
  DLM(MC) (triangles up),
 cone(MC) (triangles down) 
and  a MF fit to the experimental
    temperature reduced magnetic moment (circles).
}  
 \end{center}
\end{figure}

\begin{figure}[thbt]
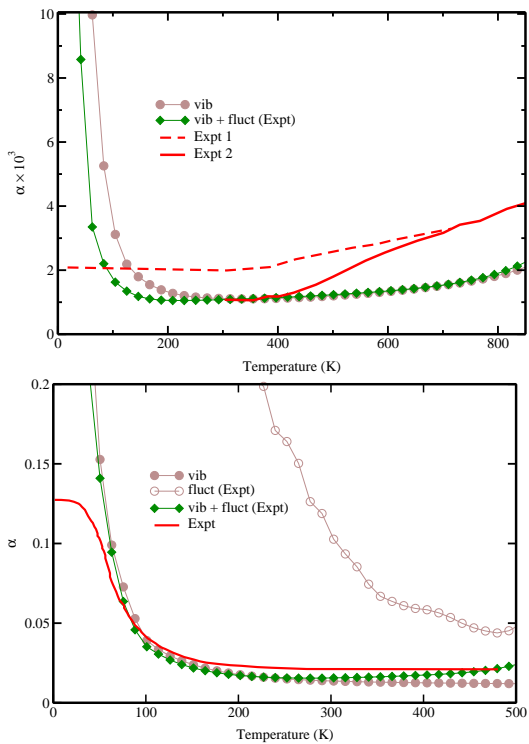

 \begin{center}
   \includegraphics[angle=0,width=0.8\linewidth,clip]{Fig_9a_CMP_Gilbert_fluct+vib_Fe_lmax3.eps}
   \includegraphics[angle=0,width=0.8\linewidth,clip]{Fig_9b_Ni_Gilbert_alfa_vs_T_vib_fluct.eps}
  \caption{\label{FIG-GIL-EXPT} 
Top:
Temperature dependent Gilbert damping
    $\alpha(T)$ for bcc-Fe, obtained by accounted for thermal vibrations
    and spin fluctuations
 accounting for 
lattice vibrations only (circles) and
    lattice vibrations and spin fluctuations based on mean-field fit to
    the experimental temperature reduced magnetic moment $M_{\rm expt} $
    (diamonds) 
compared with experimental data 
 (dashed and full lines).\cite{BL74,HF66}
Bottom: corresponding results 
for fcc-Ni.
Experimental data have been taken from Ref.\ \onlinecite{BL74}.
}
 \end{center}
\end{figure}

Fig.\ \ref{FIG-GIL-THEORY} shows results for Gilbert damping parameter $\alpha$ of bcc-Fe 
obtained using different models for the spin fluctuations.
All curves show the typical conductivity-like behavior
for low temperatures and the resistivity-like
behavior at high temperatures
reflecting the change from dominating
intra- to inter-band transitions.\cite{GIS07}
%
%
The curve denoted expt is based on a
spin configuration toted to the experimental 
$ M_{ \rm expt}(T) $ data.
Using the conical model to fit $ M_{ \rm expt}(T) $
as basis for the calculation of $ \alpha (T)$
leads obviously to a rather good agreement with
$ \alpha_{ M(\rm expt) }(T)$.
Having instead a DLM-like representation of
$ M_{ \rm expt}(T) $, on the other hand,
transverse spin components are suppressed and noteworthy
deviations from $ \alpha_{ M(\rm expt) }(T)$
are found for the low temperature regime.
Nevertheless, the deviations
are less pronounced than in the case of the longitudinal
resistivity (see Fig.\ \ref{FIG-RHO-DISP-FLUCT} (bottom)), where corresponding
results are shown based on $ M_{ \rm MC}(T) $
as a reference.
Obviously, the damping parameter $ \alpha $ seems to be less sensitive to the specific spin fluctuation model 
used than the resistivity.
Finally, using the spin configuration deduced from Monte Carlo simulations, i.e.\ 
based on $ M_{ \rm MC}(T) $
quite strong deviations for the resulting
 $ \alpha_{ M(\rm MC) }(T)$
from  $ \alpha_{ M(\rm expt) }(T)$ 
are found.
As for the resistivity (see Fig.\ \ref{FIG-RHO-DISP-FLUCT} (bottom))
this seems to reflect the too fast drop of the
reduced magnetic moment $ M_{ \rm MC}(T) $
 with temperature
in the low temperature regime compared
with temperature
(see Fig.\ \ref{FIG-MAG-T}).
As found before \cite{EMKK11}
accounting only for thermal vibrations
  $ \alpha_{ }(T)$ (Fig.\ \ref{FIG-RHO-DISP-FLUCT} (bottom))
is found comparable to the case when only
thermal span fluctuations are allowed.
Combing both thermal effects does not lead
to a curve that is just the sum of the two
 $ \alpha_{ }(T)$ curves.
As found for
 the conductivity (Fig.\ \ref{FIG-RHO-DISP-FLUCT} (top)) 
obviously the two thermal
effects are not simply additive.
As Fig.\ \ref{FIG-GIL-EXPT} (top) shows, the resulting damping parameter
 $ \alpha_{ }(T)$ 
for bcc-Fe
 that accounts for thermal vibrations as well as
spin fluctuations is found in reasonable good 
agreement with experimental data.\cite{EMKK11}
%
%
%

 Fig.\ \ref{FIG-GIL-EXPT} 
shows also corresponding results for the 
Gilbert damping of fcc-Ni as a function of temperature.
Accounting only for thermal spin fluctuations on the basis of the experimental
 $M(T)$-curve leads in this case to   completely unrealistic results
while accounting only for thermal displacements leads to results already
in rather good agreement with experiment. Taking finally both sources of disorder
into account  again no simple additive behavior 
is found but the results are nearly unchanged 
compared to those based on the thermal displacements alone.
This implies that results for the Gilbert damping parameter of fcc-Ni
hardly depend on the specific spin configuration model used but are much
more governed by thermal displacements.

\section{Summary} 
Various schemes
based on the alloy analogy that allow to include  
thermal effects 
when calculating
response properties relevant in spintronics
have been presented and discussed. 
Technical details
of an implementation
within the framework of the spin-polarized
relativistic KKR-CPA band structure method
have been outlined that allow to
 deal with thermal
vibrations as well as spin fluctuations.
Various models to represent spin fluctuations
have been compared with each other
concerning corresponding results for the
temperature dependence of the reduced magnetic moment
$M (T)$ as well as response quantities.
It was found that response quantities 
are much more sensitive to the spin fluctuation
model as the reduced magnetic moment
$M (T)$.
Furthermore, it was found that the influence
of thermal 
vibrations and spin fluctuations
is not additive when calculating electrical conductivity
or the Gilbert damping parameter $\alpha $.
Using experimental data for the reduced magnetic moment
$M (T)$ to set up
 realistic temperature dependent
spin configurations satisfying agreement for
the electrical conductivity
as well as the Gilbert damping parameter 
could be obtained for elemental ferromagnets
bcc-Fe  and fcc-Ni.

\begin{acknowledgments}
  This    work    was   supported    financially    by   the    Deutsche
  Forschungsgemeinschaft (DFG) 
within the projects EB154/20-1, EB154/21-1 and  EB154/23-1 
as well as 
the 
priority program SPP 1538 (Spin Caloric Transport)
 and the 
SFB 689 (Spinph\"anomene in reduzierten Dimensionen).
Helpful discussions with Josef Kudrnovsk\'y and Ilja Turek
are gratefully acknowledged.
\end{acknowledgments}


%

\end{document}